\documentclass[pre,a4paper,twocolumn,final,hyperref,nofootinbib]{revtex4}
\usepackage{amsfonts}
\usepackage{amsmath}
\usepackage{amssymb}
\usepackage{graphicx}

\begin{document}

\title[]{On superstatistical
multiplicative-noise processes}
\author{S\'{\i}lvio M. Duarte Queir\'{o}s}
\altaffiliation{Present address: \textit{Unilever R \& D, Port Sunlight, Quarry Road East , 
Bebington, Wirral CH63 3JW, United Kingdom}}
\email{sdqueiro@cbpf.br, sdqueiro@googlemail.com}
\affiliation{Centro Brasileiro de Pesquisas F\'{\i}sicas, Rua Dr. Xavier Sigaud, 150,
22250-250 Rio de Janeiro - RJ, Brasil}
\pacs{PACS number}

\begin{abstract}
In this article we analyse the long-term probability density function of
non-stationary dynamical processes with time varying multiplicative noise exponents 
which are enclosed inwards the Feller class of processes. The update
in the value of the exponent occurs in the same conditions as presented by \textsc{Beck} and 
\textsc{Cohen} for superstatistics. Moreover, we are able to provide
a dynamical scenario for the emergence of a generalisation of the Weibull
distribution previously introduced.
\end{abstract}

\date{\today }
\maketitle

\section{Introduction}

The description within a physical context of driven non-equilibrium complex
systems has frequently been made by considering that their dynamical
behaviour is characterised by spatial-temporal fluctuations of some
parameter, $\tilde{\beta}$. Usually, this parameter has been considered to
be the (inverse) temperature, the dissipation of energy in turbulent flows,
the amplitude of Gaussian white noise, the \textit{local} mean-reverting
value or the \textit{local} variance. As an example, we mention the standard
case of a Brownian particle diffusing along an inhomogeneous medium in which
temperature (hence diffusion \textquotedblleft constant\textquotedblright)
fluctuates in both space and time. In this approach, as it can be
understood, there are two important time scales: the scale in which the
dynamics is able to reach a stationary state (assuming a fixed value for
parameter $\tilde{\beta}$), and the scale at which the fluctuating parameter
evolves. A particular case to consider is when these two time scales are
clearly separated, specifically, when the time needed for the system to
reach stationarity (considering a predetermined $\tilde{\beta}$) is much
smaller than the scale at which that parameter changes. In the long-term,
the non-equilibrium system is described by the superposition of different
local dynamics at different time intervals that was coined by \textsc{Beck%
} and \textsc{Cohen} as \textit{superstatistics} or
\textquotedblleft statistics of statistics\textquotedblright\ \cite%
{beck-cohen,a-b-c}. Frequently, systems that are characterised as
\textquotedblleft superstatistical\textquotedblright\ exhibit non-Gaussian
distributions with kurtosis excess, or distributions with non-exponential
decay. In addition, superstatistical systems present a parameter, $\tilde{%
\beta}$, that fluctuates on a large scale, $T$, and follows a
time-independent distribution, $p(\tilde{\beta})$. The superstatistical
framework has successfully been applied on a widespread of problems like:
interactions between hadrons from cosmic rays \cite{wilk}, fluid turbulence~%
\cite{beck-prl,b-c-s,rizzo}, granular material~\cite{beck-physa},
electronics~\cite{sattin}, economics~\cite%
{economics,arch,volumes-epl,volumes}, among many others~\cite{mix}.
Furthermore, it has been regarded as a possible foundation for non-extensive
statistical mechanics~\cite{beck-prl} based on Tsallis entropy~\cite{ct} as
we show later on. In this manuscript we introduce a different analysis of
differential stochastic dynamics in which the exponent of a Feller process
is assumed as the superstatistical parameter. The main advantage of this
proposal is that it permits the evolution of the functional form of the
second-order Kramers-Moyal moment in opposition to previous presentations.

\section{Superstatistics}

Consider an inhomogeneous system composed by a large set of cells
that have different values of some parameter $\tilde{\beta}$ as we have
referred to here above. Within each cell, local equilibrium is reached very
promptly. The parameter $\tilde{\beta}$ is taken as constant throughout a
period of time $T$ after which it changes into a new value. This update
occurs always in accordance with a distribution $p(\tilde{\beta})$ for the
parameter. Taking into consideration that each cell is in local equilibrium,
thus presenting a Boltzmann factor, $e^{-\beta \,E}$,\footnote{$E$ is the
effective energy in each cell.} the long-term stationary distribution~%
\footnote{%
The observation time $t\gg T$.} of the non-equilibrium system is obtained
from a weighted average of \textit{local} Boltzmann factors, with $\beta
\equiv \beta \left( \tilde{\beta}\right) $, 
\begin{equation}
P\left( E\right) =\int p\left( \tilde{\beta}\right) \,\rho \left( E\right) \,%
\frac{e^{-\beta E}}{Z\left( \beta \right) }\,d\tilde{\beta},
\end{equation}%
where $\rho \left( E\right) $ represents the density of states, and $Z\left(
\beta \right) $ the normalisation constant. Going back to the example of a
Brownian particle moving across a medium treated in Ref.~\cite{beck-prl}, we
get that its velocity, $\vec{v}$, is obtained from the local Langevin
equation,%
\begin{equation}
d\,\vec{v}=-\gamma \,\vec{v}\,dt+\sigma \,d\vec{W}_{t}.  \label{langevin}
\end{equation}%
Seeing that the medium is inhomogeneous, either $\gamma $ \cite%
{talkner-hanggi} or $\sigma $ vary from cell to cell on a large time scale $%
T $ \footnote{%
If $\gamma $ is the random parameter then, $\tilde{\beta}=\gamma $, else if $%
\sigma $ is the random parameter then, $\tilde{\beta}=\sigma .$}. Therefore,
local Boltzmann factor has, 
\begin{equation}
\beta = \frac{2\,\gamma }{m\,\sigma ^{2}},  \label{ausloos}
\end{equation}%
which is random. From Eq. (\ref{ausloos}), the parameter $\beta $ can also
fluctuate for the case of a particle with varying mass~\cite{ausloos-massa}.

Within time scale $T$, and according to Eq.~(\ref{langevin}), the local
stationary distribution of velocities is a Gaussian conditioned to value $%
\beta$,%
\begin{equation}
p^{\prime}\left( \vec{v}\,|\beta\right) =\left( \frac{\beta}{2\,\pi }\right)
^{d/2}\exp\left[ -\frac{1}{2}\beta\,m\,\vec{v}^{2}\right] .  \label{maxwell}
\end{equation}

If the system is able to reach some local equilibrium before an update of $%
\tilde{\beta}$ takes place, \textit{i.e.}, $T\gg\gamma^{-1}=\tau$, then, we
can determine the marginal velocities probability distribution of the
long-term behaviour of the Brownian particle,%
\begin{equation}
P\left( \vec{v}\right) =\int_{0}^{\infty}p\left( \beta\right) \,p^{\prime
}\left( \vec{v}\,|\beta\right) \,d\beta.  \label{p-v}
\end{equation}
Hence, it is straightforward to verify that the form of $P\left( \vec
{v}%
\right) $ depends explicitly on the functional form of $p\left( \beta\right) 
$. Specifically, it was verified in Ref.~\cite{beck-prl} that, when $p\left(
\beta\right) $ is the $\chi^{2}-$distribution with $n$ degrees of freedom,
Eq. (\ref{p-v}) yields,%
\begin{equation}
P\left( \vec{v}\right) =\frac{1}{Z}\left[ 1+\left( 1-q\right) \,\beta _{0}\,%
\vec{v}^{2}\,\right] ^{1/\left( 1-q\right) },  \label{p-v-final}
\end{equation}
where $q=1+\frac{2}{n+d}$, $Z$ is the normalisation factor, and $\beta_{0}$
the average inverse temperature (see Ref.~\cite{beck-cohen} for details).
Such a distribution $P\left( \vec{v}\right) $ maximises Tsallis entropy \cite{ct}, 
\begin{equation}
S_{q}=\frac{1-\int\left[ p\left( x\right) \right] ^{q}\,dx}{q-1}\qquad\left(
q\in\Re\right) .  \label{s-q}
\end{equation}
This fact has turned out superstatistics into the first dynamical scenario
for the emergence of non-extensive statistical mechanics~\cite{cohen}.

\section{The Model}

Consider the following one-dimensional stochastic differential equation,%
\begin{equation}
dv=-\gamma v\,dt+\omega \left[ v^{2}\right] ^{\alpha }\,dW_{t},\qquad \left(
v\neq 0\quad \mathrm{if}\quad \alpha <0\right) ,  \label{sde}
\end{equation}%
where $W_{t}$ is a regular Wiener stochastic process, \textit{i.e.}, $%
\left\langle dW_{t}\right\rangle =0$, and $\left\langle
dW_{t}\,dW_{t^{\prime }}\right\rangle =dt\,\delta \left( t-t^{\prime
}\right) $ \footnote{%
The main advantage of writing $\left[ v^{2}\right] ^{\alpha }$ instead of $%
\left\vert v\right\vert ^{\alpha ^{\prime }}$ with \mbox{$\alpha ^{\prime
}=2\alpha $} is that of analyticity for all $v$ when $\alpha >0$.}. 
Stochastic equation (\ref{sde}) belongs to the Feller class of
(multiplicative noise) processes~\cite{feller} with $\gamma \geq 0$, $\alpha
<\frac{1}{4}$ for a (time-dependent) normalisable probability density
function (PDF) $f\left( v,t\right) $. The associated Fokker-Plank Equation
of Eq. (\ref{sde}) is%
\begin{equation}
\frac{\partial f\left( v,t\right) }{\partial t}=\frac{\partial }{\partial v}%
\left[ \gamma \,v\,f\left( v,t\right) \right] +\frac{1}{2}\frac{\partial ^{2}%
}{\partial v^{2}}\left[ \omega ^{2}\left[ v^{2}\right] ^{2\alpha }\,f\left(
v,t\right) \right] ,
\end{equation}%
whose solution $f\left( v,t\right) $ relaxes exponentially with a
characteristic time, $\tau $, into the stationary
solution,%
\begin{equation}
p\left( v\right) =\frac{1}{Z}\exp \left[ -\frac{\gamma }{\omega ^{2}\left(
1-2\,\alpha \right) }v^{2\left( 1-2\alpha \right) }\right] \left(
v^{2}\right) ^{-2\alpha },  \label{pv}
\end{equation}%
\textit{i.e.}, a Weibull-like distribution, $\mathcal{W}\left( v\right) $. 
$Z $ is the normalisation constant, $\int p\left( v\right) \ dv$,%
\begin{equation}
Z=\frac{2}{1-4\alpha }\left[ \frac{\gamma }{\omega ^{2}\left( 1-2\alpha
\right) }\right] ^{\frac{1-4\,\alpha }{4\,\alpha -2}}\Gamma \left[ 2+\frac{1%
}{4\,\alpha -2}\right] .
\end{equation}%
For $\alpha =0$, Eq. (\ref{sde}) becomes the standard Langevin equation, and 
$p\left( v\right) $ the Gaussian distribution, 
\begin{equation}
\mathcal{G}\left( v\right) =\frac{1}{Z}\exp \left[ -\frac{\gamma }{\omega
^{2}}v^{2}\right] ,  \label{gauss}
\end{equation}%
with $Z=\sqrt{\frac{\pi \,\omega ^{2}}{\gamma }}$.

\begin{figure}[tbh]
\begin{center}
\includegraphics[width=0.75\columnwidth,angle=0]{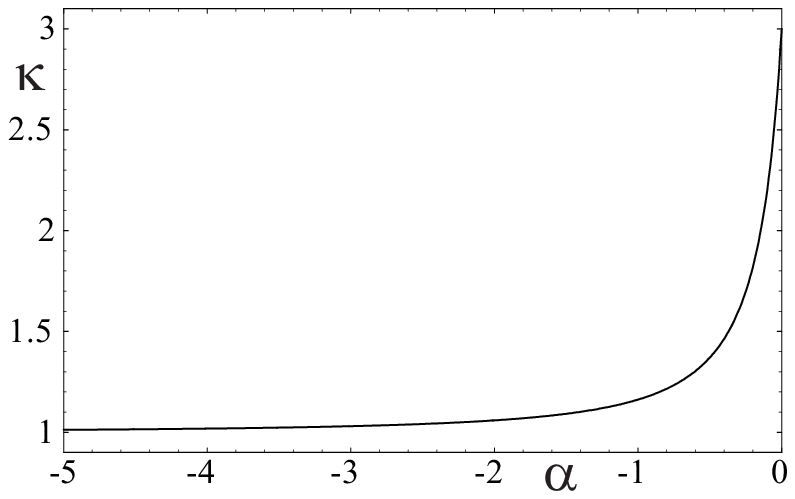} %
\includegraphics[width=0.75\columnwidth,angle=0]{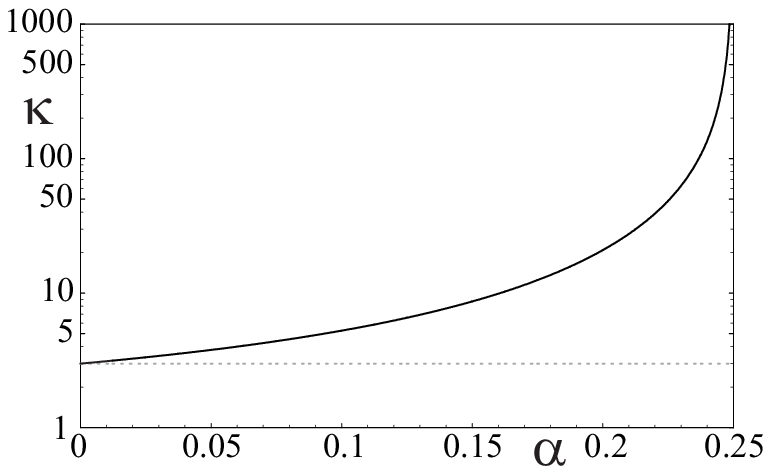}
\end{center}
\caption{ Kurtosis, $\protect\kappa $, vs. $\protect\alpha $. Upper panel:
Non-positive values of $\protect\alpha $ which lead into platykurtic
distributions. In the limit $\protect\alpha \rightarrow -\infty $, $\protect%
\kappa =1$. Lower panel: Non-negative values of $\protect\alpha $ leading
into leptokurtic distributions which has $\protect\alpha =\frac{1}{4}$ as
upper bound. The dashed line corresponds to the kurtosis of a Gaussian $\kappa =3$.}
\label{fig-05}
\end{figure}

After the transient, $f\left( v,t\right) \approx p\left( v\right) $. The
mean value of $v$, $\bar{v}\equiv \int v\,p\left( v\right) \,dv$, is equal
to zero as well as all odd moments of $p\left( v\right) $. Regarding the
second-order moment, $\overline{v^{2}}\equiv \int v^{2}\,p\left( v\right)
\,dv$, we have got%
\begin{equation}
\overline{v^{2}}=\left\{ 
\begin{array}{ccc}
\frac{\omega ^{2}}{2\gamma } & \mathrm{if} & \alpha =0 \\ 
&  &  \\ 
\frac{4\alpha -1}{4\alpha -3}\left[ \frac{\omega ^{2}\left( 1-2\alpha
\right) }{\gamma }\right] ^{\frac{1}{1-2\,\alpha }}\frac{\Gamma \left[ \frac{%
5-8\,\alpha }{2-4\,\alpha }\right] }{\Gamma \left[ 2+\frac{1}{4\,\alpha -2}%
\right] } & \mathrm{if} & \alpha \neq 0%
\end{array}%
\right. .  \label{variance}
\end{equation}%
If we consider $v$ as the velocity of a particle of unitary mass which does
a $1D$ random walk, using equipartition theorem we are able to
determine the inverse temperature, $\beta \equiv \left( k\,T\right) ^{-1}$~\footnote{ 
In this expression $T$ is the tempetature.},
yielding%
\begin{equation}
\beta =\overline{v^{2}}.  \label{equiparticao}
\end{equation}%
Evaluating the kurtosis%
\begin{equation}
\kappa \equiv \left( \overline{v^{2}}\right) ^{-2} \int
v^{4}\,p\left( v\right) \,dv,  \label{kurtosisdef}
\end{equation}%
we have obtained%
\begin{equation}
\kappa =\left\{ 
\begin{array}{ccc}
3 & \mathrm{if} & \alpha =0 \\ 
&  &  \\ 
\frac{\left( 3-4\alpha \right) ^{2}}{\left( 4\alpha -5\right) \left( 4\alpha
-1\right) }\frac{\Gamma \left[ \frac{7-8\,\alpha }{2-4\,\alpha }\right]
\,\Gamma \left[ 2+\frac{1}{4\,\alpha -2}\right] }{\left\{ \Gamma \left[ 
\frac{5-8\,\alpha }{2-4\,\alpha }\right] \right\} ^{2}} & \mathrm{if} & 
\alpha \neq 0%
\end{array}%
\right. .
\end{equation}%
As it is visible from Fig. \ref{fig-05},\ distribution (\ref{pv}) is
platykurtic for $\alpha <0$, and leptokurtic for $\alpha >0$.

Moving on, we shall now consider that \emph{exponent }$\alpha $\emph{\
instead of constant, varies according to superstatistical requirements}. In
other words, let us consider an inhomogeneous system which is composed by a
large set of cells that have different values of $\alpha $. Within each
cell, the value of $\alpha $ is updated at every elapsed time interval $\ell 
$, in agreement with a certain PDF $\rho \left( \alpha
\right) $. The update scale, $\ell $, is much greater than relaxation time
scale $\tau $. Noticing that each cell is in local equilibrium, the
long-term stationary distribution of the non-equilibrium system is obtained
performing the integral,%
\begin{equation}
P\left( v\right) =\int \rho \left( \alpha \right) \,p\left( v\right)
\,d\alpha .
\end{equation}%
As possible applications of such a model we name: description of velocities in
granular material (particularly see figures of Ref.~\cite{welles}) and unconventional turbulent fluids, or even the dynamics of
financial observables. Explicitly, systems which undergo through different phases during their
time evolution or situations where different stages are measured when
observations are made at the same point of space. In a thermodynamic context, the
fluctuations in $\alpha $ correspond to fluctuations in temperature, but
obtained through a completely different way from the proposal presented in Ref.~\cite%
{beck-prl}. Specifically, fluctuations in $\alpha $ induce a modification of
the functional form of the $2^{nd}$ order Kramers-Moyal coefficient, while
in Ref.~\cite{beck-prl} its functional form is always preserved.

\subsection{Some examples}

\subsubsection{Dichotomous case}

This case represents the simplest form to introduce fluctuations in $\alpha $%
, and for which a full analytical treatment is possible. Its probability
density function is simply,%
\begin{equation}
\rho \left( \alpha \right) =\frac{1}{2}\delta \left( \alpha -\alpha
_{0}\right) +\frac{1}{2}\delta \left( \alpha -\alpha _{1}\right) ,\qquad
\left( \alpha _{0}\neq \alpha _{1}\right) .  \label{dichotomous}
\end{equation}%
Amongst all endless possibilities for $\alpha _{0}$ and $\alpha _{1}$, let
us firstly consider cases for which one of the exponents is equal to zero.
The long-term distribution is thus given by%
\begin{equation}
P\left( v\right) =\frac{1}{2}\mathcal{G}\left( x\right) +\frac{1}{2}\mathcal{%
W}_{\alpha }\left( x\right) .  \label{dirac}
\end{equation}%
For small values of $\left\vert v\right\vert $, and $\alpha <0$, $P\left(
v\right) $ approaches the limit $\sqrt{\frac{\gamma }{4\,\pi \,\omega ^{2}}}$ as $%
\left[ v^{2}\right] ^{-2\alpha }$.

\begin{figure}[tbh]
\begin{center}
\includegraphics[width=0.75\columnwidth,angle=0]{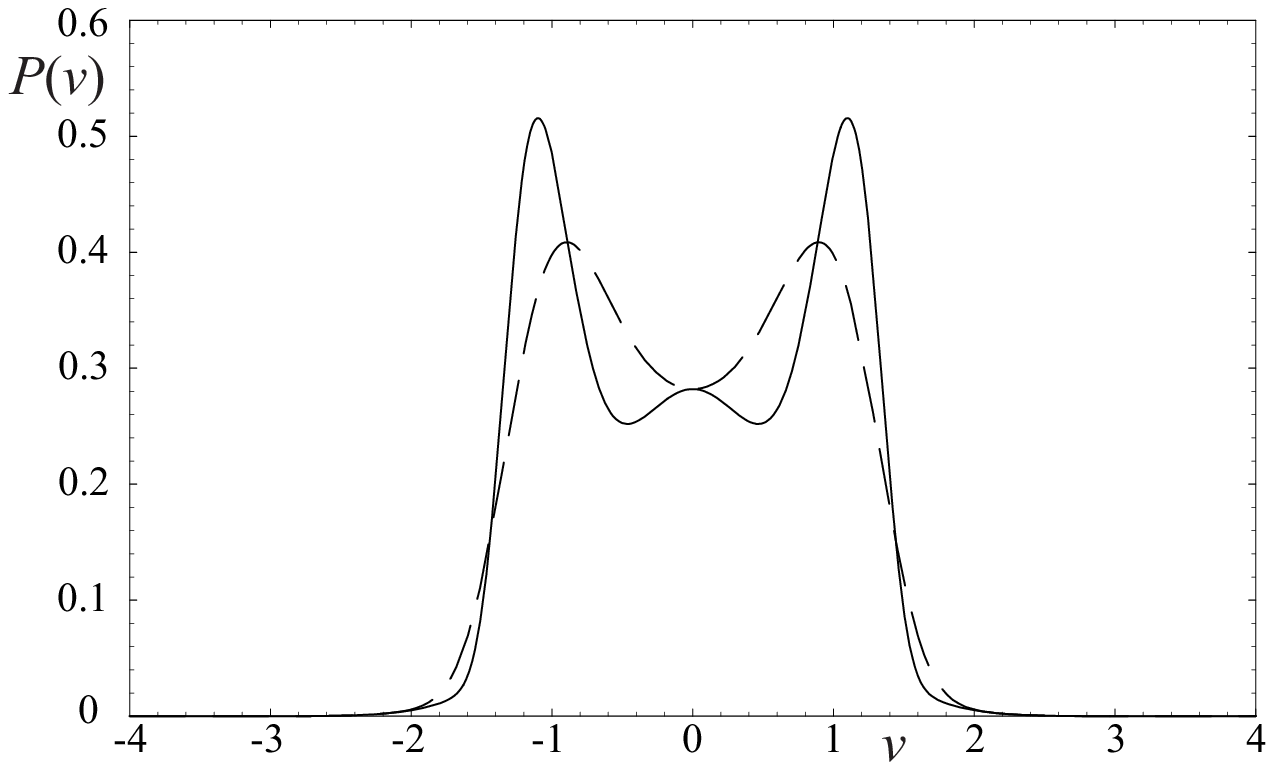} %
\includegraphics[width=0.75\columnwidth,angle=0]{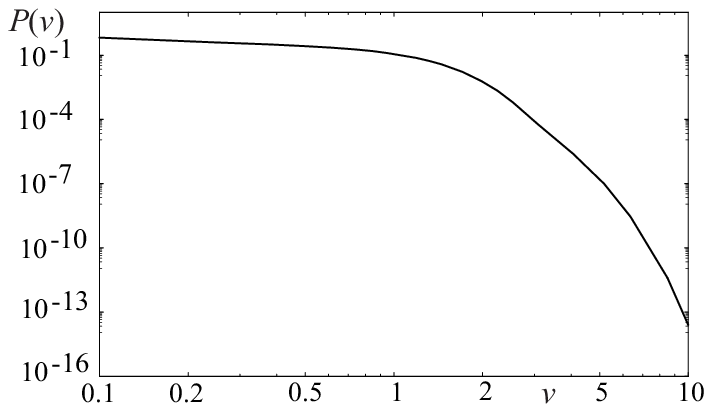}
\end{center}
\caption{ Probability density function $P\left( v\right) $ \textit{vs.} $v$
for the case of superstatistical processes with a dichotomous distribution (%
\protect\ref{dichotomous}) where $\protect\gamma =\protect\omega =1$. Upper
panel: In both cases $\protect\alpha _{0}=0$, and $\protect\alpha _{1}=-1$
for the full line, and $\protect\alpha _{1}=-\frac{1}{2}$ for the dashed
line. Lower panel: $P\left( v\right) $ vs. $v$ for a dichotomous case with $%
\protect\alpha _{0}=0$ and $\protect\alpha =\frac{1}{5}$ ($\protect\gamma =%
\protect\omega =1$) in a log-log scale. For this case, the variance $\overline{v_{ef}^{2}}%
=0.2860\ldots $ and kurtosis $\protect\kappa =11.953\ldots $. }
\label{fig-1}
\end{figure}

\begin{figure}[tbh]
\begin{center}
\includegraphics[width=0.75\columnwidth,angle=0]{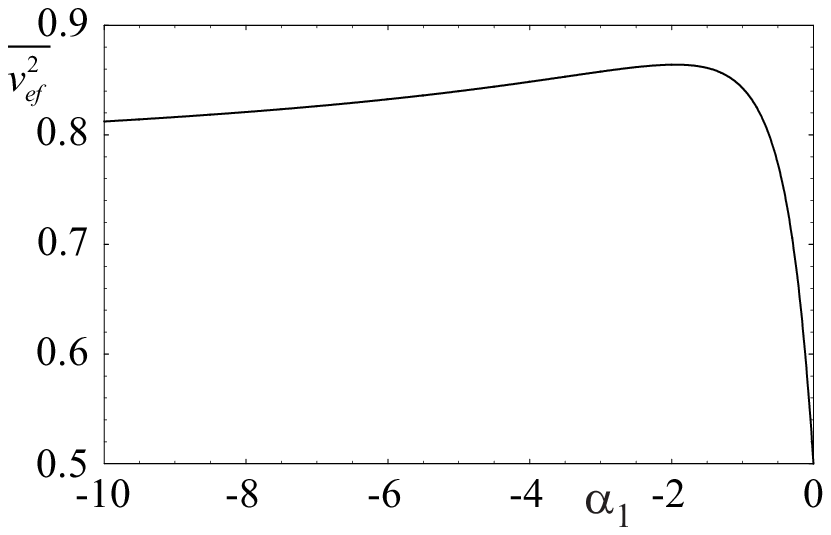} %
\includegraphics[width=0.75\columnwidth,angle=0]{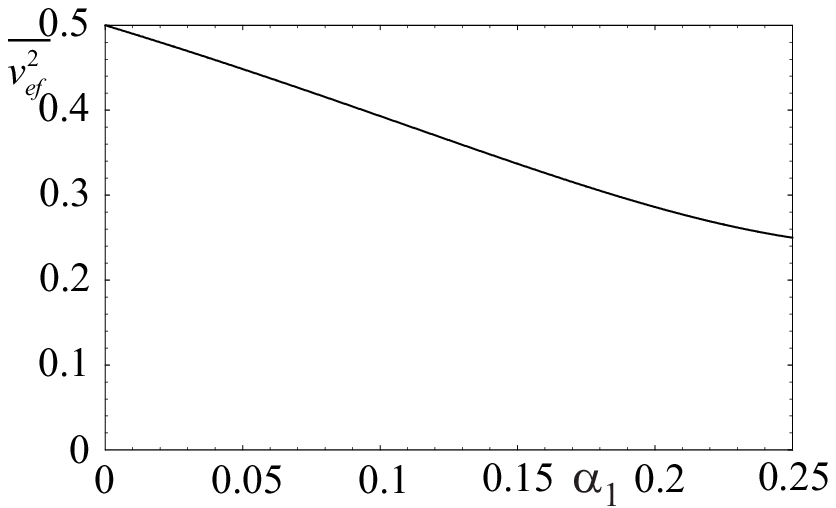}
\end{center}
\caption{ Standard deviation, $\overline{v_{ef}^{2}}$, \textit{vs.}
dynamical exponent $\protect\alpha _{1}$ for superstatistical process with a
dichotomous PDF (\protect\ref{dichotomous}) where $\protect\alpha _{0}=0$,
and $\protect\gamma =\protect\omega =1$. For $\protect\alpha _{1}=0$, $%
\overline{v_{ef}^{2}}=\frac{1}{2}$, and for $\protect\alpha _{1} =-\infty $ the
value of $\overline{v_{ef}^{2}}$ tends to $\frac{3}{4}$ (upper panel). For
positive values of $\protect\alpha _{1}$, $\overline{v_{ef}^{2}}$ is a
monotonically decreasing function of $\protect\alpha _{1}$ with $\overline{%
v_{ef}^{2}}=\frac{1}{4}$ for $\protect\alpha _{1}=\frac{1}{4}$.}
\label{fig-2}
\end{figure}

In Fig.~\ref{fig-1} we exhibit the resulting probability density function, $%
P\left( v\right) $, for $\alpha _{0}=0$, and $\alpha _{1}=-1$, $\alpha _{1}=-%
\frac{1}{2}$, and $\alpha =\frac{1}{5}$. The value $\overline{v_{ef}^{2}}$,%
\begin{equation}
\overline{v_{ef}^{2}}\equiv \int v^{2}\,P\left( v\right) \,dv,
\label{temperaturadirac}
\end{equation}%
is presented in Fig.~\ref{fig-2} for several values of $\alpha _{1}$.

\subsubsection{Uniform distribution case}

\begin{figure}[tbh]
\begin{center}
\includegraphics[width=0.75\columnwidth,angle=0]{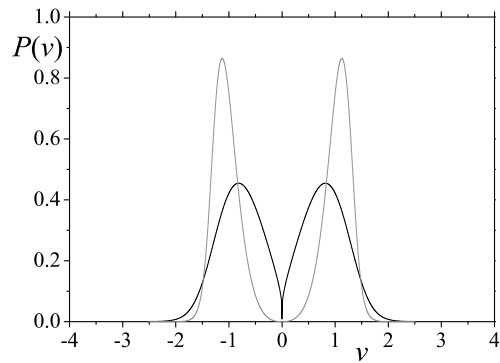}
\end{center}
\caption{ Numerically obtained probability density function $P\left(
v\right) $ \textit{vs.} $v$ for the case of superstatistical processes with
uniform distribution, Eq.~(\ref{uniform}) where $\protect\gamma =%
\protect\omega =1$. The black line corresponds to $\protect\alpha _{0}=0$,
and $\protect\alpha _{1}=-\frac{1}{2}$, with $\overline{v_{ef}^{2}}%
=0.833\ldots $, and kurtosis, $\protect\kappa =1.755\ldots $. The grey line
corresponds to a uniform distribution with $\protect\alpha _{0}=-\frac{1}{2}$
$\protect\alpha _{1}=-\frac{3}{2}$ yielding $\overline{v_{ef}^{2}}%
=1.169\ldots $, and kurtosis, $\protect\kappa =1.179\ldots $. }
\label{fig-3}
\end{figure}

If, for the previous example the analytical form of $P\left( v\right) $ is
easily obtained, that does not happen in situations in which $\alpha $
evolves according to a uniform distribution between $\alpha _{0}$ e $\alpha
_{1}$, 
\begin{equation}
\rho \left( \alpha \right) =\left\{ 
\begin{array}{ccc}
\frac{1}{\alpha _{0}-\alpha _{1}} & \mathrm{if} & \alpha _{1}\leq \alpha
\leq \alpha _{0} \\ 
&  &  \\ 
0 & \mathrm{otherwise} & 
\end{array}%
\right. .  \label{uniform}
\end{equation}%
However, it is possible to evaluate numerically the form of the long-term
distribution as we present for two particular cases in Fig.~\ref{fig-3}.
Computing the kurtosis, $\kappa $, for both cases we have verified that the
two examples are platykurtic.

\begin{figure*}[tbh]
\begin{center}
\includegraphics[width=0.9\columnwidth,angle=0]{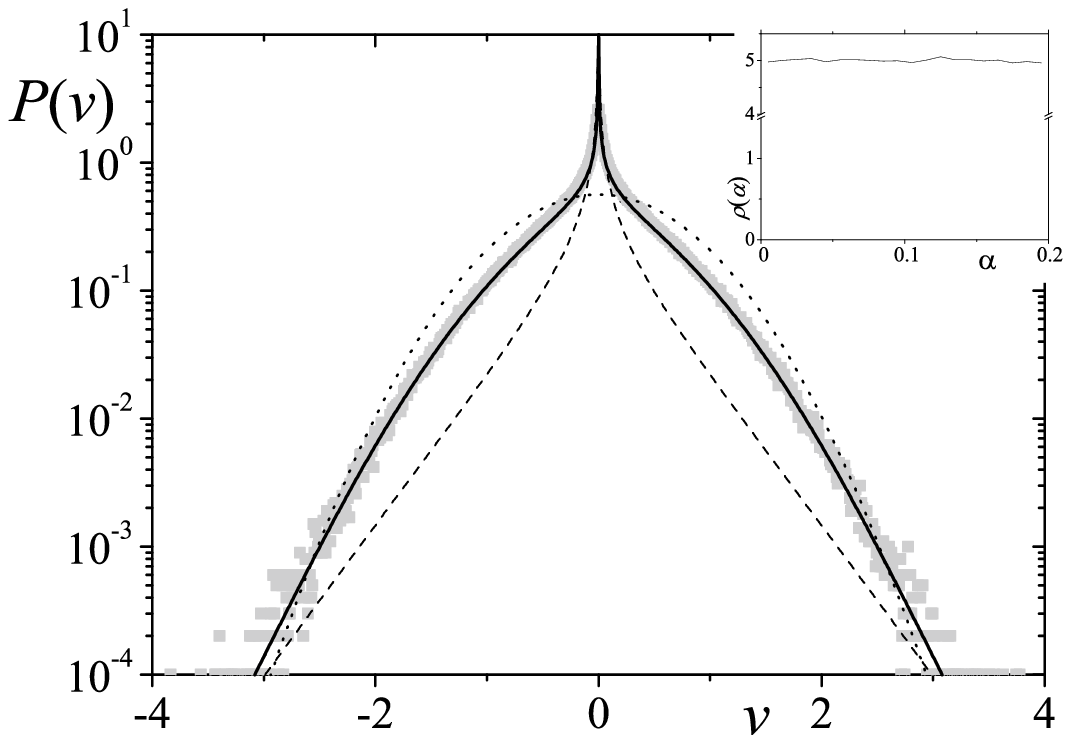} %
\includegraphics[width=0.9\columnwidth,angle=0]{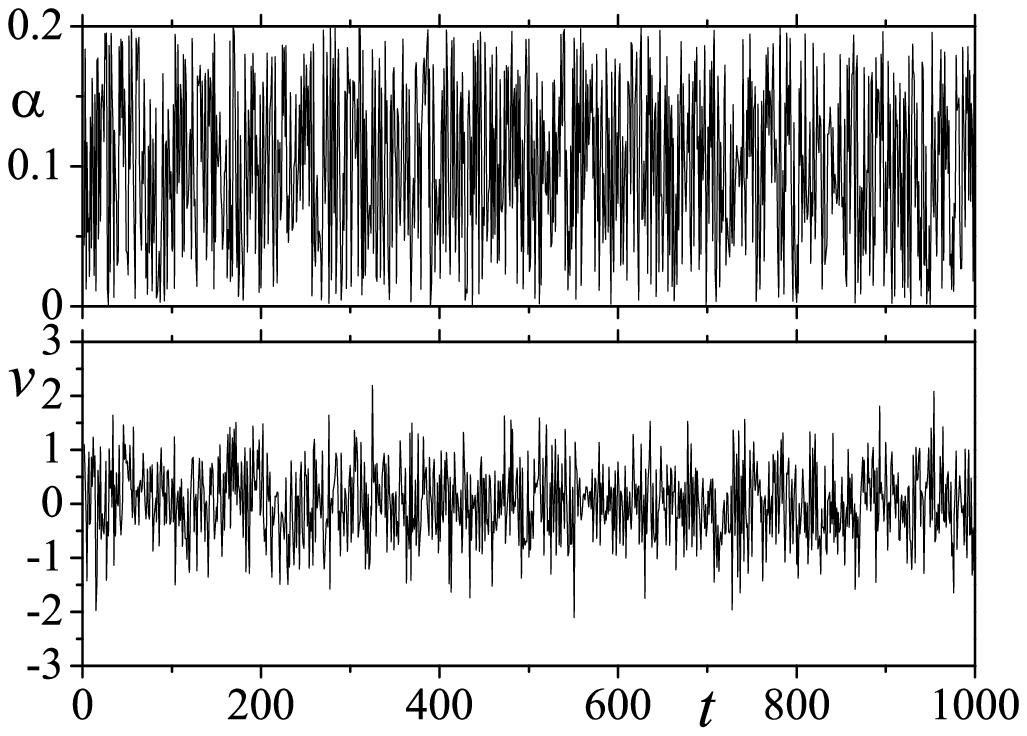} %
\end{center}
\caption{
Left panel: The full line represents the numerically obtained probability
density function $P\left( v\right) $ \textit{vs.} $v$ for the case of
superstatistical processes with uniform distribution, Eq. (\ref{uniform}),
where $\alpha _{0}=0$, and $\alpha _{1}=\frac{1}{5}$ $\left( \gamma
=100,\omega =10\right) $. The dashed line represents the PDF of the upper
bound $\alpha =\frac{1}{5}$, and the dotted line the PDF of the lower bound $%
\alpha =0$ (a Gaussian). The long-term PDF decays slower than the Gaussian.
The grey symbols represent the PDF obtained from the numerical simulation of
a superstatistical system with the same parameters $\gamma $ and $\omega $
and PDF $\rho \left( \alpha \right) $. In the inset we show $\rho \left( \alpha \right) $ of
that process. Right panels:\ Excerpt of a superstatistical time series of $v$
which evolves according to Eq. (\ref{sde}) with $\gamma =100$, $\omega =10$,
and $\alpha $ associated with a uniform distribution between $\alpha =0$ and 
$\alpha =\frac{1}{5}$. The time scale of updating $\alpha $ is $1$ time unit
which is rather larger than $10^{-2}$ that is the time scale of relaxation
towards stationarity. For this case $\alpha _{0}=0$, and $\alpha _{1}=\frac{1}{5}$, with $\overline{%
v_{ef}^{2}}=0.285\ldots $, and kurtosis, $\kappa =5.272\ldots $. A fair similar distribution, namely Fig. 6(b), 
has been obtained in Ref.~\cite{catania-tsallis} for the velocities PDF of a long-range Hamiltonian system
at a quasi-stationary state.
}
\label{fig-31}
\end{figure*}

Another possibility is to consider just non-negative values for $\alpha $. 
This situation is more likely to be experimentally verified,
since kurtosis excess is quite ubiquitous. In this case, and since we are
considering leptokurtic distributions, $P\left( v\right) $ is also
leptokurtic. An illustration of this sort of example is presented in Fig.~\ref{fig-31} 
where $\alpha $ uniformly varies between $0$ and $\frac{1}{5}$.

\begin{figure}[tbh]
\begin{center}
\includegraphics[width=0.8\columnwidth,angle=0]{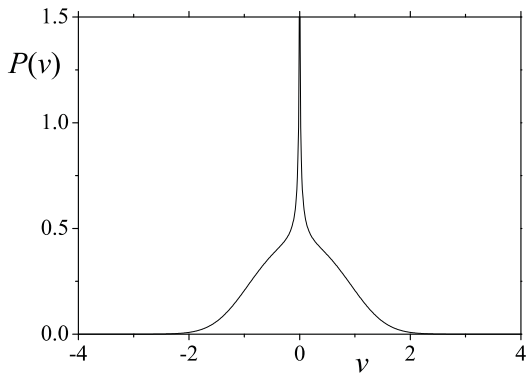}
\end{center}
\caption{
Numerically obtained probability density function $P\left( v\right) $ 
\textit{vs.} $v$ for the case of superstatistical process with uniform
distribution, Eq. (\ref{uniform}), where $\alpha _{0}=-\frac{1}{5}$, and $%
\alpha _{1}=\frac{1}{5}$ $\left( \gamma =\omega =1\right) $. In this case $%
\overline{v_{ef}^{2}}=0.476\ldots $, and kurtosis, $\kappa =3.129\ldots $.
Interestingly, this example suggests the existence of, at least, one
distribution $\rho \left( \alpha \right) $ with $\alpha _{0}\neq \alpha
_{1}\neq 0$ for which $P\left( v\right) $ is mesokurtic.}
\label{fig-32}
\end{figure}

Moreover, extending our range of values for exponent $\alpha $, we can
consider positive and negative values as we present in Fig.~\ref{fig-32}.
For this last case, in the long-term, the system endures both platykurtic
and leptokurtic regimes. 

\subsubsection{The $\protect\chi^{2}$-distribution}

\begin{figure}[tbh]
\begin{center}
\includegraphics[width=0.75\columnwidth,angle=0]{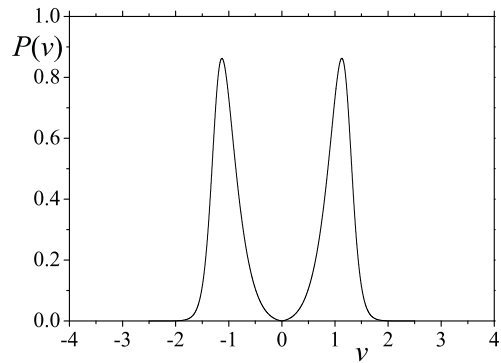}
\end{center}
\caption{ Numerically obtained probability density function $P\left(
v\right) $ \textit{vs.} $v$ for the case of superstatistical processes with
a $\protect\chi^{2}$-distribution, Eq. (\protect\ref{chi2}) where $\protect%
\nu=5$, and $\overline{\alpha }=\protect\gamma =\protect\omega=1$. For this
example we have $\overline{v_{ef}^{2}}=1.150\ldots$, and kurtosis, $\protect%
\kappa=1.999$. }
\label{fig-4}
\end{figure}

Another distribution that appears in various phenomena is the $\chi ^{2}$%
-distribution, 
\begin{equation}
\rho \left( \alpha \right) =\frac{\nu ^{\nu }}{\bar{\alpha}\,\Gamma \left[
\nu \right] }\left( \frac{\left\vert \alpha \right\vert }{\bar{\alpha}}%
\right) ^{\nu -1}\exp \left[ -\frac{\nu }{\bar{\alpha}}\left\vert \alpha
\right\vert \right] .  \label{chi2}
\end{equation}%
For this case an analytical form is not, in principle, possible to obtain.
Nonetheless, we obtain the numerical solution for $P\left( v\right) $ as we
show in Fig.~\ref{fig-4}.

\subsection{A generalised Weibull distribution}

\begin{figure*}[tbh]
\begin{center}
\includegraphics[width=0.75\columnwidth,angle=0]{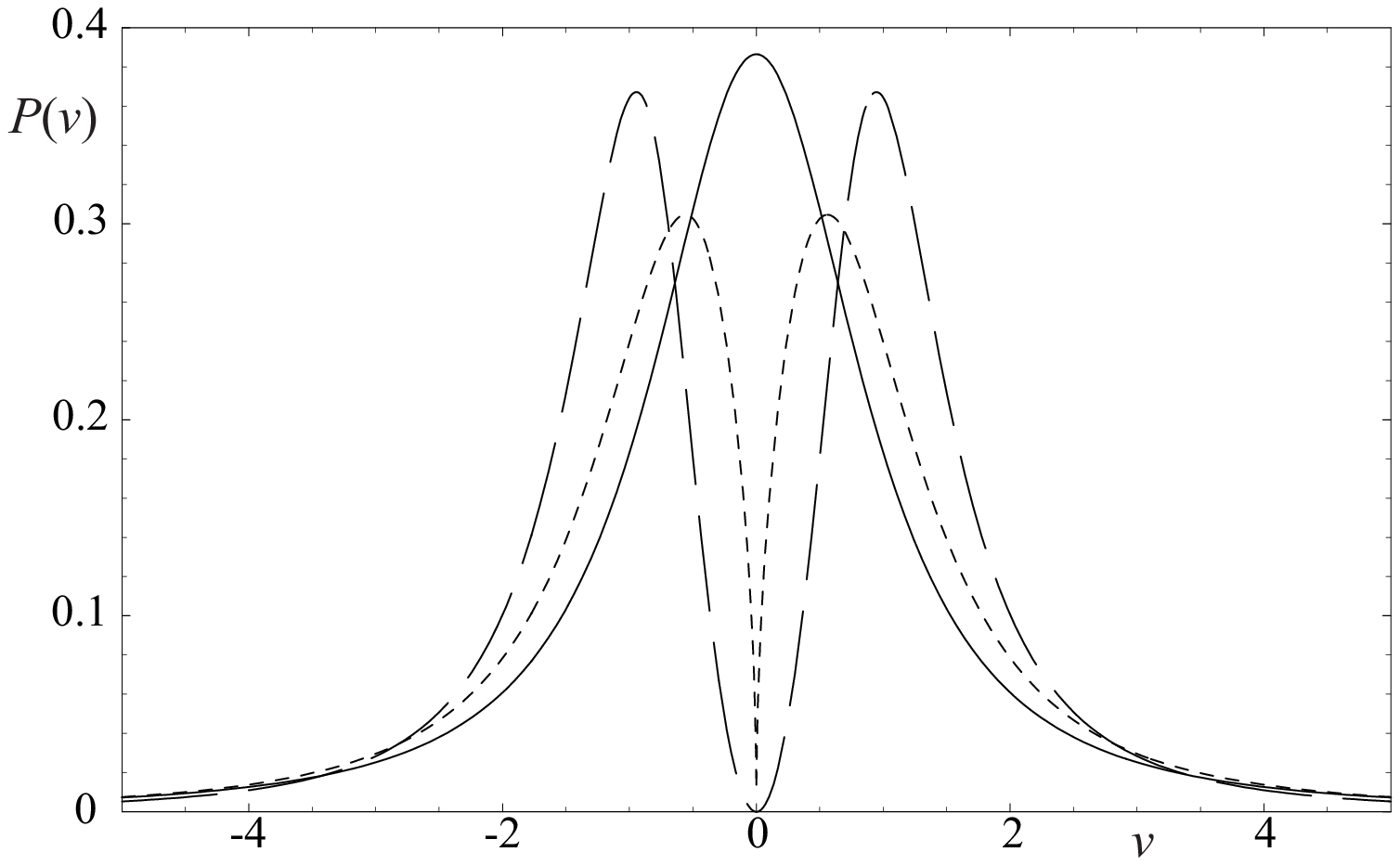} %
\includegraphics[width=0.75\columnwidth,angle=0]{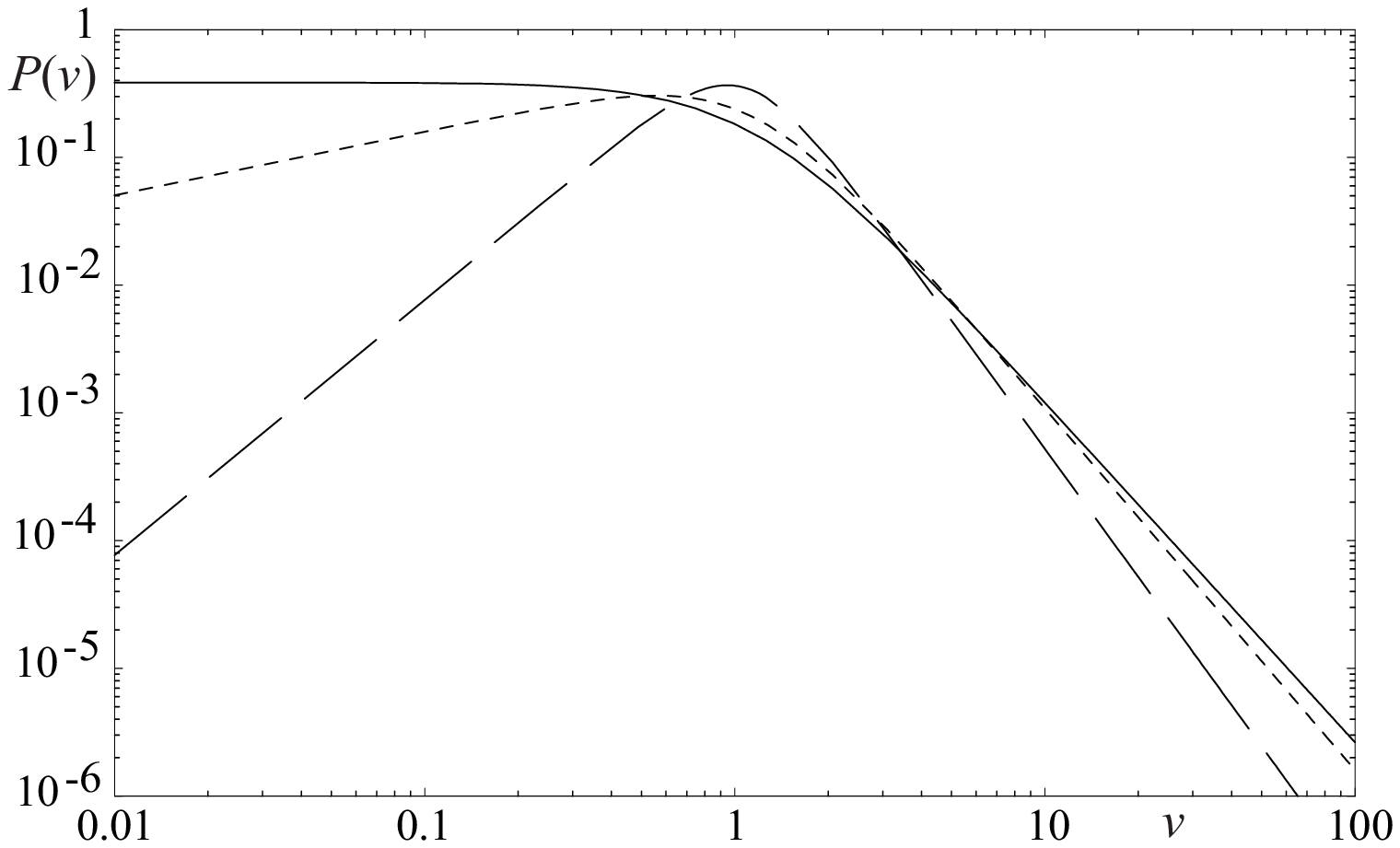} %
\includegraphics[width=0.75\columnwidth,angle=0]{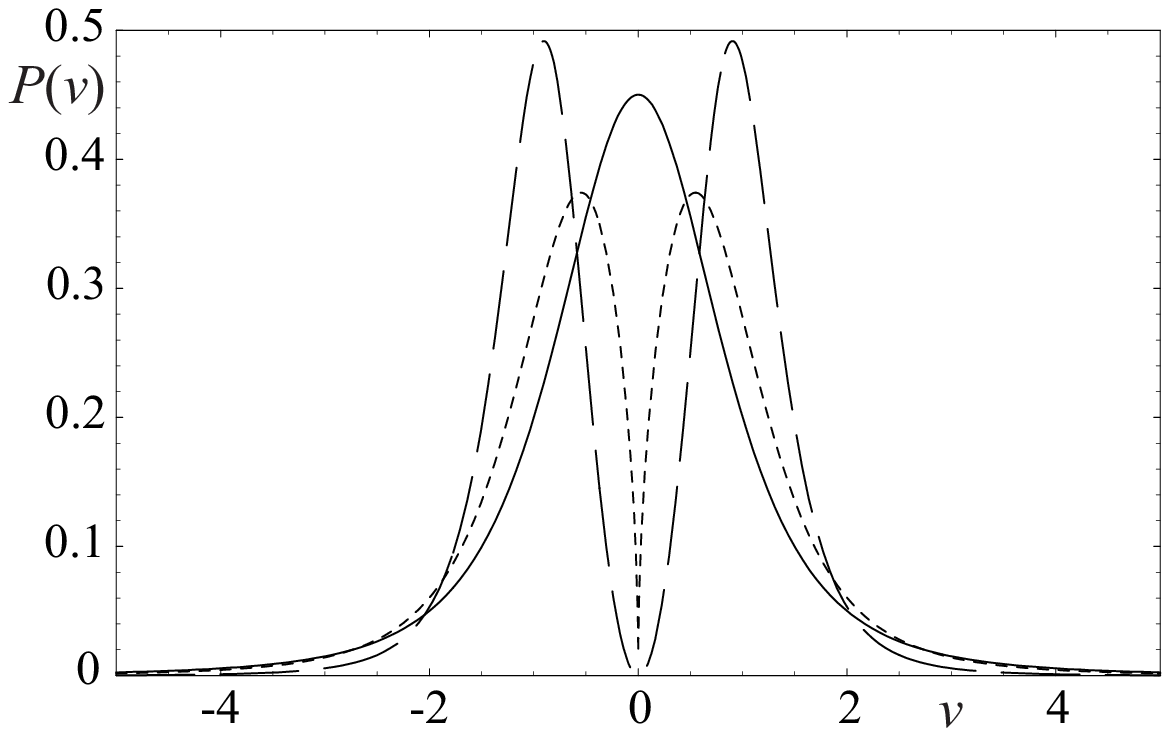} %
\includegraphics[width=0.75\columnwidth,angle=0]{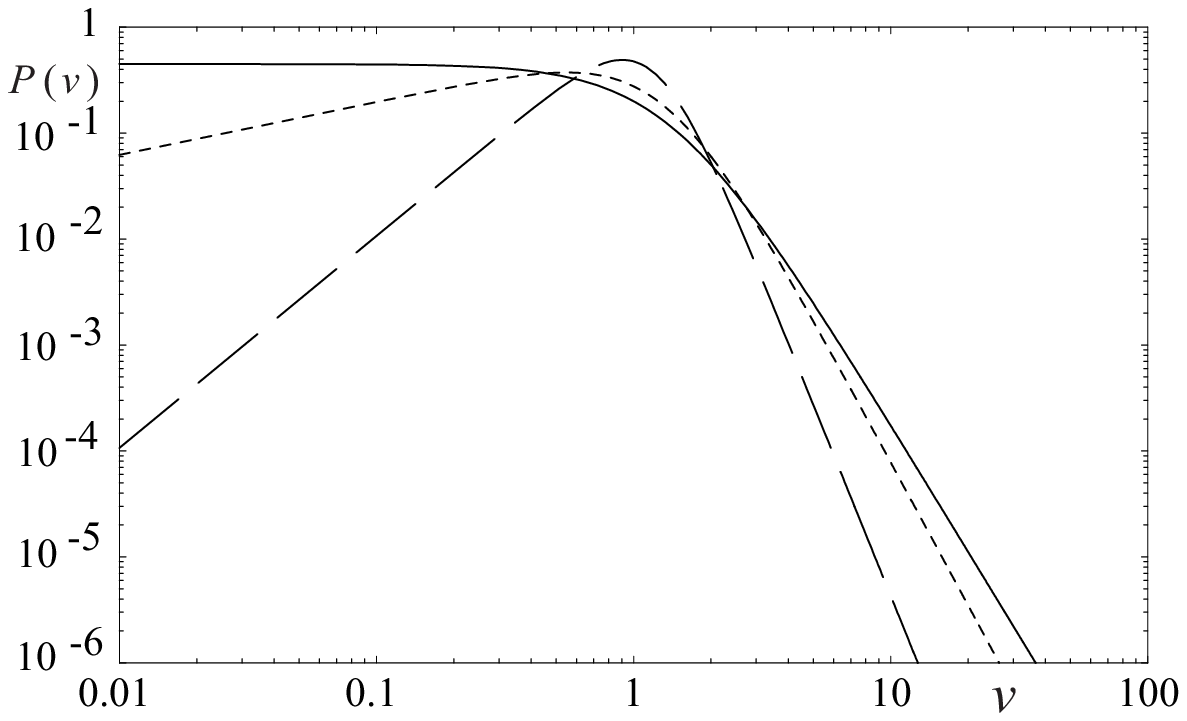} %
\includegraphics[width=0.75\columnwidth,angle=0]{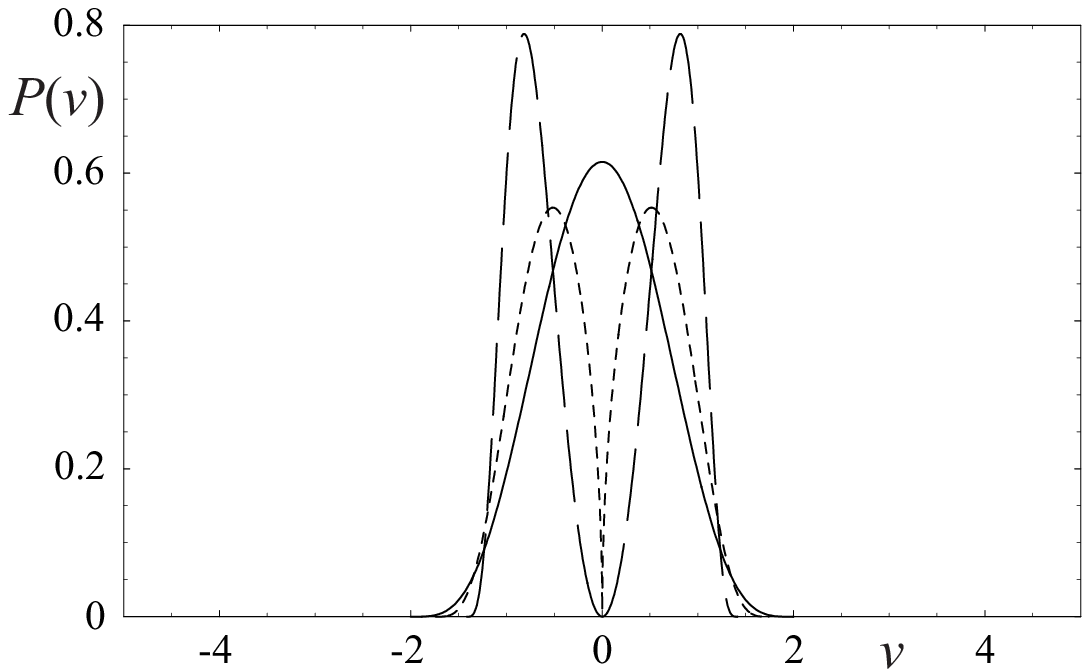} %
\includegraphics[width=0.75\columnwidth,angle=0]{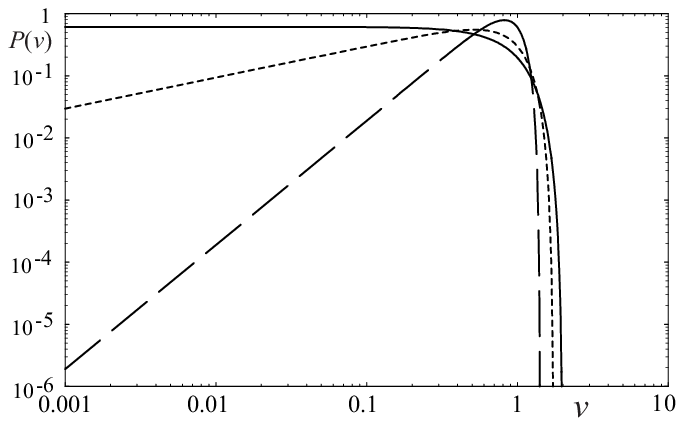}
\end{center}
\caption{ Representation of $_{q}\mathcal{W}\left( v\right) $ \textit{vs.} $%
v $ for several values of parameters $a,b,q$ , and $\tilde{v}=1$. Left
panels contain linear-linear representations, whereas right panels exhibit
the same $_{q}\mathcal{W}\left( v\right) $ but in a $\log $-$\log $ scale.
For long dashed lines $\left( a=2,b=1\right) $; short dashed lines $\left( a=%
\frac{5}{4},b=\frac{1}{4}\right) $; and for the full lines $\left(
a=1,b=0\right) $ which correspond to a $q$-Gaussian PDF. In upper panels we
have $q=\frac{7}{4}$. Because of conditions (\protect\ref{condition}), the $%
\left( a=2,b=1\right) $ case is the only one to have a finite $\overline{%
v_{ef}^{2}} $ . Explicitly, $\overline{v_{ef}^{2}}=6.424\ldots $. Midst
panels: $q=\frac{3}{2}$. Both of the three cases have a finite $\overline{%
v_{ef}^{2}}$. Specifically, $\overline{v_{ef}^{2}}=\protect\sqrt{2}$ for $%
\left( a=2,b=1\right) $, $\overline{v_{ef}^{2}}=2$ for $\left(
a=1,b=0\right) $, and $\overline{v_{ef}^{2}}=2^{4/5}$ for $\left( a=\frac{5}{%
4},b=\frac{1}{4}\right) $. Curiously, case $\left( a=2,b=1\right) $ has $%
\protect\kappa =3$ just like a Gaussian distribution, \textit{i.e.}, a
mesokurtic distribution. Lower panels: $q=\frac{3}{4}$. In this case,
distributions present a compact support, $\left[ -v_{c},v_{c}\right] $, by
reason of Tsallis cut-off, where $v_{c}=\left( \frac{1}{1-q}\right) ^{\frac{1%
}{2\,a}}$. }
\label{fig-5}
\end{figure*}

In this subsection we treat, the standard case where in Eq. (\ref{sde}) $%
\omega $ evolves on a superstatistical fashion rather than $\alpha $. For
this particular case, when $\Omega \equiv \omega ^{-2}$ follows a $\chi ^{2}$%
-distribution,%
\begin{equation}
\rho \left( \Omega \right) =\frac{1}{\Omega _{0}\,\Gamma \left[ \frac{\nu }{2%
}\right] }\left( \frac{\nu }{2\,\Omega _{0}}\right) ^{\frac{\nu }{2}}\Omega
^{2+\frac{\nu }{2}}\exp \left[ -\frac{\nu }{2}\frac{\Omega }{\Omega _{0}}%
\right] ,
\end{equation}%
the long-term stationary distribution, $P\left( v\right) $, that corresponds
to a weighted average of $p\left( v\right) $ over all possible values of $%
\Omega $, yields%
\begin{equation}
_{q}\mathcal{W}\left( v\right) =\frac{1}{Z^{\prime }}\exp _{q}\left[ -\frac{%
\left( v^{2}\right) ^{a}}{\tilde{v}}\right] \left( v^{2}\right) ^{b},
\label{qweibull}
\end{equation}%
where,%
\begin{equation}
q=\frac{4+\nu -6\,\alpha -2\,\nu \,\alpha }{2+\nu -2\,\alpha -2\,\nu
\,\alpha },\quad \tilde{v}=\frac{\nu \left( 1-2\,\alpha \right) ^{2}\Omega
_{0}}{\left( 2+\nu -2\,\alpha \left( 1+\nu \right) \right) \gamma },
\end{equation}%
and%
\begin{equation}
a=1-2\,\alpha ,\quad b=-2\,\alpha .
\end{equation}%
Owing to its comparability to the Weibull distribution inside $S_{q}$
framework, and following Ref.~\cite{renio-qweibull} we shall call PDF (\ref%
{qweibull}) $q$\emph{-Weibull distribution}. The $q$-Weibull distribution
has been used to numerically adjust a variety of PDFs, but, to the best of
my knowledge, no dynamical basis has been presented so far. Distribution (\ref%
{qweibull}) belongs to the Burr class of probability density functions~\cite%
{burr}. For small values of $\left\vert v\right\vert $, $_{q}\mathcal{W}\left( v\right) $ in Eq.~(\ref%
{qweibull}) goes to zero as a power law with exponent $b$, and for large $%
\left\vert v\right\vert $, the same distribution also vanishes as a power
law but with exponent $a/\left( 1-q\right) +b$. For this case, it is possible
to evaluate even moments of order $m$ when the following conditions are
verified,%
\begin{equation}
a>0,\quad b>-\frac{m+1}{2},\quad \frac{a}{q-1}-b>\frac{m+1}{2}.
\label{condition}
\end{equation}
When $b=0$, \textit{i.e.}, $\alpha =0$, $P\left( v\right) $ turns into a $q$%
-Gaussian distribution recovering the scenario of Ref.~\cite{beck-prl}. In Fig.~%
\ref{fig-5} we depict examples of $_{q}\mathcal{W}\left( v\right) $ for some
values of $a$, $b$, and $q$.

\section{Remarks and perspectives}

In this manuscript we have discussed a superstatistical approach of a
stochastic process which belongs to the Feller class, and whose (local)
stationary probability density function is reminiscent of a Weibull
distribution. Such proposal, which might be seen has a dynamical analogue for Ref.~\cite{q-super}, has been driven on 
two ways. The first one in which the power of multiplicative noise term is time-dependent, and another
one that is equivalent to the evolution of the noise width in the stochastic term of the corresponding differential equation.
The main advantage of the former is the possibility of
mimicking systems which hold a rather complex dynamics that goes through a
(random) sequence of dynamical regimes. We have determined either
analytically or numerically the long-term probability density function which
can assume all types of kurtosis. Concerning the latter superstatistical
approach, we have verified that it allows the emergence of a generalisation
of the Weibull distribution within the framework of Tsallis non-additive
entropy. This distribution has proved to be valid for numerical adjustments
in a large variety of systems. For the $q$-Weibull, we have also been able
to obtain platykurtic, mesokurtic, and leptokurtic distributions. In respect
of is actual implementation to the analysis of experimental data we must
refer that it requires the development of techniques to capture both of $%
\rho \left( \alpha \right) $ and $T$ as it was introduced for cases with 
fluctuations in $\omega $~\cite{b-c-s}\cite{smdq-supercrit}. The solution
for this non-trivial challenge either on an experimental or theoretical
level is certainly welcomed.

Last of all, the extension of the idea of a stochastic dynamics with superstatistical
exponents that lead to a consistent emergence of leptokurtic long-term
probability density functions by means of considering, \textit{e.g.}, a
non-zero mean or the combination of independent multiplicative and additive
noises will be the focus of further research.

\begin{acknowledgments}
I am deeply grateful to Professor Constantino Tsallis who has always stimulated my
scientific curiosity with his observations, comments, and endless incentive. 
I thank D. O. Soares-Pinto for calling my attention to the work of Ref.~\cite{welles}, and to
E.M.F Curado and F.D. Nobre for their ever useful comments.
This work benefited from financial support from Funda\c{c}\~{a}o para a Ci%
\^{e}ncia e Tecnologia (Portuguese Agency), and also infrastructural support
from PRONEX (Brazilian agency). The present piece of work pays homage to Brasil and its 
flamboyant people where and with whom I have had the best times of my life.
\end{acknowledgments}

\medskip
{\it Note after acceptance} - During a talk given in November $2007$ at Queen Mary University 
of London based on this work I took knowledge by Prof. C. Beck of an article on turbulence, 
namely, Ref.~\cite{castaing}, in which are presented probability density functions with an aspect very similar 
to some of the distributions presented herein. This underlines the surmised application of the our theoretical 
model to turbulence models.

\end{document}